\documentclass[structabstract]{aa}

\usepackage{graphicx}
\usepackage{amssymb}
\usepackage{amsmath}

\usepackage{natbib}

\usepackage{txfonts}

\begin{document}

\title{A cosmological independent calibration of the $E_{p,i}$-$E_{iso}$ correlation for Gamma Ray Bursts}

\author{S. Capozziello\inst{1}
\and L. Izzo\inst{1,2,3} }

\institute{Dipartimento di Scienze Fisiche, Universit\`a di Napoli
"Federico II" and INFN Sez. di Napoli, Compl. Univ. Monte S.
Angelo, Ed. N, Via Cinthia, I-80126 Napoli, Italy \and ICRANet and
ICRA, Piazzale della Repubblica 10, I-65122 Pescara, Italy. \and
Dip. di Fisica, Universit\`a di Roma "La Sapienza", Piazzale Aldo
Moro 5, I-00185 Roma, Italy.}

\abstract {}{The relation connecting  the emitted
isotropic energy  and the rest-frame peak energy of the $\nu$F$\nu$ spectra of Gamma-Ray Bursts (the Amati relation),
strictly depends on the cosmological model, so we need a method to obtain an independent calibration of it.}
{Using the Union Supernovae Ia catalog, we obtain a cosmographic luminosity distance in the $y$-redshift and demonstrate that this parametrization approximates very well the fiducial standard comsomlogical model $\Lambda$CDM. Furthermore, by this cosmographic luminosity distance $d_l$, it is possible to achieve the Amati relation independent  on the cosmological model}
{The cosmographic Amati relation that we obtain agrees, in the errors,  with other cosmological-independent calibrations proposed in the literature.}{ This could be considered a good indication in view to obtain standard candles by Gamma-Ray Bursts}


\maketitle

\section{Introduction}

Supernovae $Ia$ (SNeIa) are considered  accurate and reliable standard candles, \citep{Phillips}.
In recent years, their use as cosmological distance indicators have led to the puzzling discovery that the Universe is in a phase of accelerated expansion, \citep{Riess,Perlmutter}.
This feature has also led to the revision of the standard cosmological model, leading to what is known today as the $\Lambda$CDM concordance model, see e.g. \citep{Ostriker}.
However  it is not possible to observe these objects very far in the Universe.
The most distant Supernova $Ia$ was observed at a redshift of z $\sim$ 1.7 \citep{Benitez}.
For this reason,  several cosmological analysis made by using the various compiled sample of SNeIa, like the Union Catalog, \citep{Kowalski}, are not able to investigate  higher  redshift regions of the Universe. If we had  distance indicators at  higher redshifts,  then we could extend our knowledge at these  unexplored regions.

One of the  possible solutions to this problem could come from the
Gamma-Ray Bursts (GRBs) assumed as cosmological indicators
\citep{Piran,Meszaros}. GRBs are the most powerful explosions in
the Universe: this feature allows them to be observed  at
extremely high redshift. The most distant GRB observed up to now
is at a redshift of $\sim$ 8.1, \citep{Tanvir,Salvaterra}.
However, GRBs are not standard candles, since they have no known
and well-defined luminosity relation. Due to this lack,   we have
to find another way to use GRBs as cosmological beacons. A
possible solution could consist in finding correlations between
photometric and/or spectroscopic properties of GRBs themselves. In
the scientific literature there are several of these relations,
\citep{Schaefer}. One of these is the Amati relation,
\citep{Amati}, which relates the isotropic energy emitted by a GRB
with the peak energy in the rest-frame of the $\nu$F($\nu$)
electromagnetic spectrum of a GRB. This relation has already been
widely used to constraining the cosmological density parameter
\citep{Amati2}, with quite remarkable results. However, there is
still not a physical link between this correlation and the
mechanisms underlying the production and the emission of a GRB.
The basic emission process of a GRB is very likely not unique, so
it is not easy to explain, from a physical point of view,  such a
relation. Recently it has been suggested that the Amati relation
could depend strongly on the satellite measurements used for
detection and the observation of each GRB \citep{Butler07}.
However this hypothesis has been rejected recently,
\citep{Amati2}, since the relation seems to be verified regardless
of the satellite considered for the observations and detection.

Although not supported by self-consistent physical motivations, it
is a phenomenological relation which could be extremely useful for
cosmological considerations. However, a problem related with such
a relation  is that it must be calibrated independently of the
considered cosmological model. In order to compute the energy
emitted from an astrophysical object at a certain redshift $z$, we
need, as a matter of fact, a measurement of the bolometric flux
and  the distance of the same object. For the first quantity, we
follow the idea outlined by \citep{Schaefer} :  one can obtain a
very precise measurement of the bolometric fluence emitted by a
GRB from the observed fluence, the integrated flux in the
observation time and the spectral model that best fits the
spectral energy distribution of each GRB. However, the distance
depends  on the considered cosmological model. People usually
adopt the standard  $\Lambda$CDM model, with fixed values of the
density parameter $\Omega_i$. This procedure leads  to the
so-called {\it circularity problem}  when the Amati relation is
used to standardize GRBs. For this reason we need a
cosmology-independent calibration of the  relation.

Recently, it was released a calibration with SNeIa data by  using
different numerical interpolation methods \citep{Liang}; the
results seem very reliable to address cosmological issues by GRBs.
In this work we shall take into account a similar analysis: by
taking into account   SNeIa data from the cosmographic point of
view  (for a detailed description see e.g.
\citep{Weinberg,Visser}),  it could be possible to obtain a
calibration of the Amati relation. We will use  results obtained
from a cosmographic fit of a sample of SNeIa extended up to very
high redshift with the GRBs. The use of the cosmography to deduce
the cosmological parameters from SNeIa was widely discussed in the
literature, \citep{Visser2}, and the results are very close to
that attained by other and more accurate analysis. Recently
applications  of cosmographic methods have taken into account
galaxy clusters \citep{Capozz}  and  GRB, \citep{Izzo,Vitagliano}
but their  reliability drastically fails  at high redshifts.
Indeed, the estimates of the deceleration parameter  $q_0$ and of
the jerk  parameter $j_0$  are usually achieved   only at very low
redshift and then any extrapolation could led to shortcomings and
misleading results as soon as they are extended. However by an
appropriate parameterization of the redshift parameter, one can
circumvent the problem   introducing  a new redshift variable
ranging from  0 and 1 \citep{Visser}. Let us consider the
following quantity as the new redshift variable:
\begin{equation}
 y = \frac{z}{1+z}\,,
\end{equation}
we obtain that the range of variation is between  0 and 1. In this
way, we can derive a  luminosity distance by which   we can obtain
the  Amati relation suitable for cosmography.

The layout of the paper is the following:  in Sect. 2 we  tackle
the cosmographic analysis considering the SNeIa Union sample.
Results will be used to derive the luminosity distance for each
GRB and then we will fit the  cosmographic Amati relation. In
Sect. 3 a discussion on how to extend the same relation  is
reported. We add further 13 GRBs (as of December 2009), computing
the bolometric fluence and the peak energy for each of them and
after we calculate the cosmographic parameters using the new
relation. Finally,  we  calculate the isotropic energy for each
GRB and then  compute the best fit for the considered sample of
data (Sect. 4). Discussion and conclusions are reported in Sec. 5.

\section{Cosmographic analysis}

The main purpose of this work  consists in obtaining an Amati
relation independent of the adopted cosmological model. All we
need is a formulation of the luminosity distance  $d_l$ as a
function of the redshift $z$. These two quantities are linked
together via the scale factor $a(t)$, which describes  the
expansion of the Universe in a Friedmann-Lemaitre-Robertson-Walker
cosmology. This means that we are  assuming only homogeneity and
isotropy but not the specific cosmological model, e.g.
$\Lambda$CDM model. It is well known that we can obtain the
function $a(t)$ from the Friedmann equations.  These equations can
be solved only if assumptions are made on  dynamics and  fluids
filling the Universe,  that is choosing a cosmological model. We
will relax  this possibility assuming only cosmography in the
sense described in \citep{Weinberg}. Since the evolution of the
luminosity distance is well known for small values of  redshift,
we can consider the power series expansion of the scale factor.
This naturally leads to an expression for the luminosity distance
in power series terms too \citep{Visser, Capozziello1}:

\begin{eqnarray}
d_l(z) =  d_H z
\Bigg\{ 1 + {1\over2}\left[1-q_0\right] {z}
-{1\over6}\left[1-q_0-3q_0^2+j_0+ \frac{k \; d_H^2}{a_0^2} \right] z^2
\nonumber
\\
{+}
{1\over24}[
2-2q_0-15q_0^2-15q_0^3+5j_0(1+2q_0)+s_0
\nonumber
\\
+ \frac{2\; k \; d_H^2 \; (1+3q_0)}{a_0^2}]\; z^3 +
\mathcal{O}(z^4) \Bigg\}
\end{eqnarray}

where $d_H = c/H_0$ and $H_0$, $q_0$, $j_0$ and $s_0$ are known as
the Hubble constant, the deceleration, the jerk and the snap
parameters respectively. In order to obtain accurate measurements
of the cosmographic parameters, we need to go up to large values
of the redshift. This goal can be achieved  by considering large
data sample as  SNeIa  \citep{Visser} and, eventually,  GRBs.

Here,  we are interested in reconstructing the relation $d_l$($z$)
by cosmographic methods in order to test correlations for GRBs .
In order to achieve this goal, we will use the data sample of
SNeIa   Union, \citep{Kowalski}  consisting of 307 supernovae up
to redshift  $z\sim 1.7$. By this data sample, it is possible to
perform a non-linear least-squares fit considering the empirical
equation given by the distance modulus obtained from the expanded
$d_l$($z$), that is:

\begin{eqnarray}
\mu(z) = 25 + {5\over\log 10}
\log \Bigg\{ d_H [ z + {1\over2} (1-q_0)z^2
\nonumber
\\
-{1\over6} (1+j_0+\frac{c^2k}{a^2H_0^2}-q_0-3q_0^2)z^3
\nonumber
\\
+ \frac{1}{24} ( 2+5j_0-2q_0-15q_0^2 +\frac{2c^2k(1+3q_0)}{a^2H_0^2}
\nonumber
\\
+10j_0q_0-15q_0^3+s_0) z^4  +
\mathcal{O}(z^4) ] \Bigg\}\,.
\end{eqnarray}

In this work we are not interested in the estimate of the cosmographic parameters but in using cosmography to constrain a GRB-energy  relation. To this aim, we will use a custom equation for the fit of the type
\begin{equation}
\mu(z) = 25 + (5/\log 10)\log (a z + b z^2 + c z^3 + d z^4)\,,
\end{equation}
 so we will compute only the parameters $a$,$b$,$c$,$d$.
Once we have an estimate of these parameters, we could easily
obtain the values of the related cosmographic parameters. To
obtain a better analysis,  we  can use a robust interpolation
method of   Levenberg-Marquardt type. The results of our data
fitting are shown in Table \ref{table:no1}.

\begin{table}
\caption{SNeIa cosmographic fit  obtained by both the redshift variables $z$ and $y$.} % title of Table
\label{table:no1} % is used to refer this table in the text
\centering % used for centering table
\begin{tabular}{c c c c} % centered columns (4 columns)
\hline\hline % inserts double horizontal lines
Parameter & value $z$-redshift & Parameter & value $y$-redshift \\ % table heading
\hline % inserts single horizontal line
$a$ & 4242 $\pm$ 176 & $a$ & 4213 $\pm$ 216  \\ % inserting body of the table
$b$ & 0.9593 $\pm$ 0.2447 & $b$ & 2.248 $\pm$ 0.863 \\
$c$ & -0.8201 $\pm$ 0.4290 & $c$ & -0.894 $\pm$ 3.576 \\
$d$ & 0.2722 $\pm$ 0.2119 & $d$ & 1.784 $\pm$ 4.057 \\
\hline %inserts single line
\end{tabular}
\end{table}

\begin{center}
\begin{figure*}
\begin{tabular}{|c|c|}
\hline
\includegraphics[height=6cm,width=8cm]{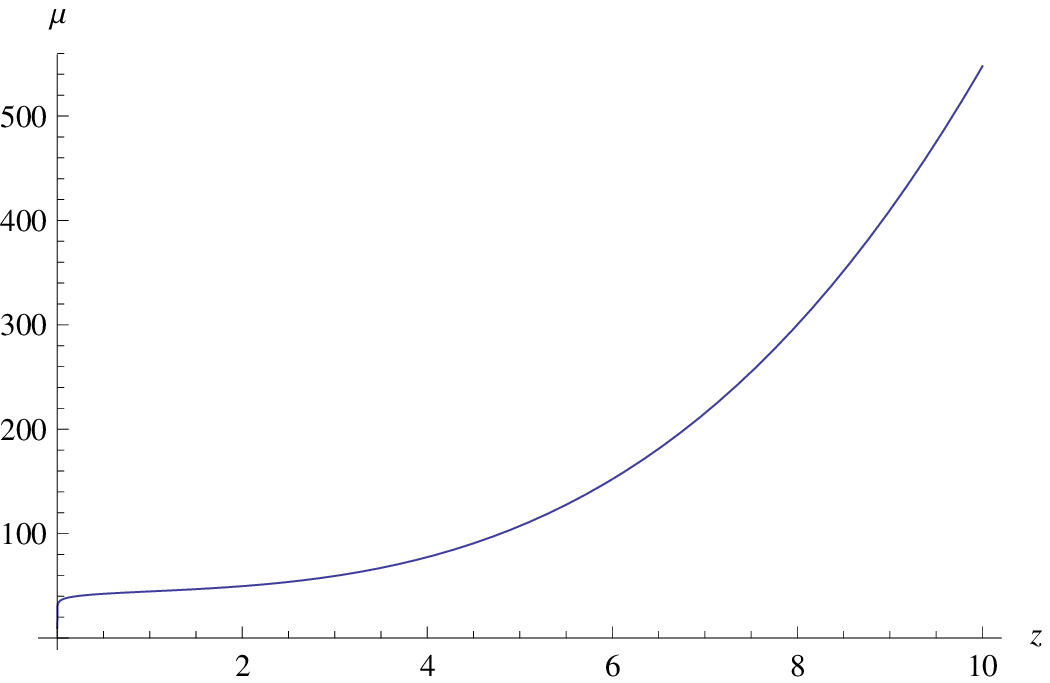}&
\includegraphics[height=6cm,width=8cm]{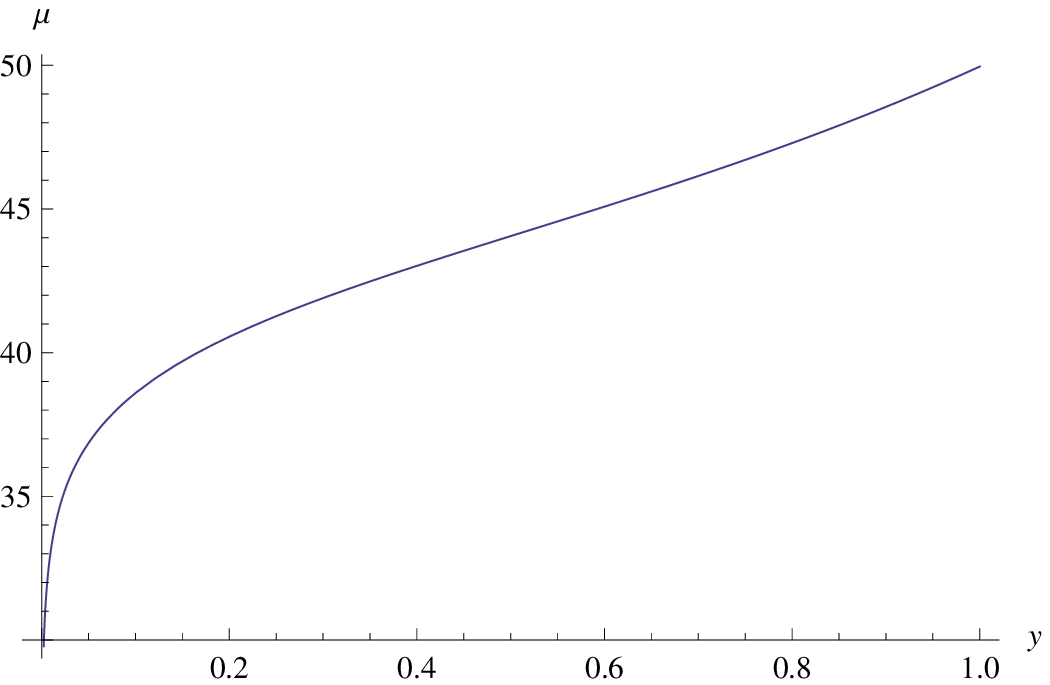}\tabularnewline
\hline
\end{tabular}
\caption{Trends of the  distance modulus for the $z$-redshift and for the $y$-redshift.} \label{fig:no1}
\end{figure*}
\end{center}

The test of  reliability of the fit has been done with a
$R^2$-test, \citep{Bevington}, whose value is $0.9914$. However
the extension up to high redshift of this function $\mu$($z$)
shows a serious problem: for redshifts greater than $\sim$ 2 the
curve grows rapidly, see Fig.\ref{fig:no1}. This steep departure
is due to the higher-order term, i.e. $d$, which has a decisive
influence at high redshift. This fact rules out {\it a priori}  a
possible supernova-calibrated $\mu$($z$) at high redshift. Such
problems can be  eliminated if we consider a new variable for the
redshift. It has been shown \citep{Visser2,Vitagliano} that the
coordinates transformation $y = z/(1+z)$ and, consequently, the
power series of the luminosity distance provides a better
extrapolation at high redshft, as well as better results for the
parameters of the fit. Due to this fact,  we can perform a
cosmographic analysis for the new distance modulus $\mu$($y$), in
analogy with what has been already done  for the $\mu$($z$). The
new expression for the distance modulus, which takes into account
the new redshift parameterization, becomes \citep{Vitagliano}:

\begin{eqnarray}
\mu(y) = 25 + {5\over\log 10}
\Bigg\{ \log d_H + \log y - {1\over2} (q_0 - 3)y
\nonumber
\\
+{1\over24} (21-4 (j_0+\frac{c^2k}{a^2H_0^2}) + q_0(9 q_0 -2)]y^2
\nonumber
\\
+ \frac{1}{24} [15 + 4\frac{c^2k}{a^2H_0^2}(q_0 -1)+j_0(8 q_0 -1)-5q_0
\nonumber
\\
+2q_0^2 -10q_0^3 +s_0]y^3 + \mathcal{O}(y^4) ] \Bigg\}
\end{eqnarray}

so we will consider a custom equation for the fit similar to the previous one, used for the estimate of the $\mu(z)$ parameters.
The results obtained with a non-linear fit are shown in Table \ref{table:no1}, while in Figure \ref{fig:no1} it is shown the trend of the distance modulus for both the redshift variables considered.

In the following we will consider the formulation for the distance modulus in terms of the $y$-redshift in order to derive a cosmographic Amati relation.

\section{The data sample}

As said in Introduction, in recent years the interest of
astrophysicists and cosmologists has been attracted by the
possibility of using GRBs as potential distance indicators. This
interest is due to the fact that most of the GRBs satisfy some
correlations between photometrical and spectroscopical observable
quantities. Among the various existing correlations (for a review
of these see e.g. \citep{Schaefer}),   the Amati relation seems
very attactive \citep{Amati}. It relates the cosmological
rest-frame $\nu$ $F(\nu)$ spectrum peak energy $E_{p,i} $ with the
equivalent isotropic radiated energy $E_{iso}$. It was discovered
based on  \emph{BeppoSAX}  data and then confirmed also for the
X-ray flashes (XRFs)  \citep{Lamb}. It seems that it does not work
for short GRBs. For this reason the  relation could be used to
discriminate among different GRB classes.

The possible origin of this correlation as due  to detector
selection effects seems not consistent, nevertheless the large
scatter in the normalization and the shift toward the Swift
detection  threshold \citep{Butler07}.  A recent study
\citep{Amati09} has  shown that the different $E_{p,i}$ - $E_{iso}
$ correlations, obtained independently from the detectors
considered for the observations, are fully consistent each other,
so the hypothesis of a strumental-dependent Amati relation seems
to fail.

Here we are going to expand the sample of GRBs reported in
\citep{Amati09} adding 13 GRBs and obtaining a sample consisting
of 108 GRBs. Substantially we need to know the redshift $z$, the
observed peak energy $E_{p,obs}$ of the $\nu$ $F(\nu)$ spectrum
and an estimate of the bolometric fluence $S_{bolo}$ for each GRB
in the sample. To derive the bolometric fluence $S_{bol}$,  we can
use the method outlined in \citep{Schaefer}, where from the
observed fluence and the spectral model, we can obtain an
estimation of $S_{bol}$ via the following formula:

\begin{equation}
S_{bol} = S_{obs} \frac{\int_{1/(1+z)}^{10^4/(1+z)} E \phi d E}{\int_{E_{min}}^{E^{max}} E \phi d E}
\end{equation}
where $\phi$ is the spectral model considered for the spectral
data fit and $S_{obs}$ is the fluence observed for each GRB in the
respective detection band  $(E_{min}, E_{max})$. In particular,
for 6 of the 13 GRBs added, we consider a cut-off power-law
spectral model while for the remaining 7 we use a band model
\citep{Band}. In the Table \ref{table:no2}, the spectral data for
the 13 GRBs are shown.  $E_p$ column refers to the measured peak
energy. To obtain the peak energy in the rest frame, we have to
take into account the redshift of the GRB, then $E_{p,i} = E_p
(1+z)$. Once we have obtained the estimate of $ S_{bol}$ for each
GRB in the sample, the next step is to estimate the isotropic
energy from the well-known formula which relates the luminosity
distance and the fluence, that is
\begin{equation}\label{eq:no4}
 E_{iso} = 4 \pi d_l^2 S_{bol} (1+z)^{-1}\,.
\end{equation}
Note that  the quantity $(1 + z)$ to obtain the  value of an
observable quantity in the rest-frame is equivalent, in the new
redshift parameterization, to use, instead, the correction $ 1 /
(1-y) $. The value of the luminosity distance which must enter in
Eq.\ref{eq:no4} is what we got previously from the cosmographic
fit of the SNeIa. From this fit, we  obtained an estimate of the
function $ \mu(y) $; to go back to the luminosity distance,  we
can use the following formula:

\begin{equation}
 d_l(y) = 10^{\frac{\mu(y)-25}{5}}
\end{equation}
by which it is possible to compute  the value of $ d_l(y)$ for each GRB in the sample.

It is worth noticing that for values of $ y $ greater than $ \sim
$ 2.5, the curve $ \mu(y)$ begins to increase slightly. This fact
could lead to improper estimates of the isotropic energies emitted
by GRBs at high redshift. If we consider an analogous curve
referred to a fiducial standard $ \Lambda$CDM cosmological model,
we can quantitatively evaluate this deviation. In Figure
\ref{fig:no3}, it is shown the deviation of the curve $ \mu(y)$,
obtained by the cosmographic fit of the SNeIa and the one obtained
by considering a $ \Lambda$CDM model with values of the density
parameters given by $ \Omega_{\rho} = 0.27$ and $\Omega_{\Lambda}
= 0.73 $. The discrepancy from the fiducial $\Lambda$CDM model
seems quite small, but it has to be taken into account when we
will compute the cosmographic Amati relation.

\begin{table*}
\centering % used for centering table
\caption{Data for the 13 GRBs added to the old sample described in \citep{Amati09}.
Table  shows: (1) the name of GRB, (2) the spectral model used for the fitting of the spectra, (3) the redshift, (4) the peak energy observed, (5) the softer spectral index, absent for the cut-off power law spectral model, (6) the higher spectral index, (7) the observed fluence and (8) the detector band considered for the estimate of the fluence, (9) the GCN reference for the GRB, where we took the spectral data.} % title of Table
\label{table:no2} % is used to refer this table in the text
\begin{tabular}{c c c c c c c c c}
\hline\hline % inserts double horizontal lines
$GRB$ & $spec$ $model$ & $z$ & $E_{p,o}$ (keV) & $\alpha$ & $\beta$ ($\gamma$) & $S_{obs}$ (10$^{-6}$ ergs/cm$^2$) & $band$ (keV) & $GCN$ \\
(1) & (2) & (3) & (4) & (5) & (6) & (7) & (8) & (9)\\
\hline % inserts single horizontal line
090516  & CPL & 4.109 & 190 $\pm$ 65    &   -- & -1.5  $\pm$ 0.3 &  15 $\pm$ 3 & 20-1200 & 9422\\ % inserting body of the table
090715B  & CPL & 3.00 & 134 $\pm$ 56  &   -- &   -1.1 $\pm$ 0.4 & 9.3 $\pm$ 1.5 & 20-2000 & 9679\\
090812  & CPL & 2.452 & 586 $\pm$ 243  &   -- &  -1.03   $\pm$ 0.07 & 26.1 $\pm$ 3.4 & 15-1400 & 9821 \\
090926B  & CPL & 1.24 & 91 $\pm$ 2  &   --    &  -0.13  $\pm$ 0.06 &  8.7 $\pm$ 0.3 & 10-1000 & 9957\\
091018  & CPL & 0.971 & 28 $\pm$ 16  &   -- &   -1.53   $\pm$ 0.59 &  1.44 $\pm$ 0.19 & 10-1000 & 10045\\
091029  & CPL & 2.752 & 61.4 $\pm$ 17.5  &   -- &   -1.46  $\pm$ 0.27 &  2.4 $\pm$ 0.1 & 15-150 & 10103\\
\hline
090618  & Band & 0.54 & 155.5  $\pm$ 11  &  -1.26 $\pm$ 0.06 &  -2.50 $\pm$ 0.33 & 270 $\pm$ 6 & 8-1000 & 9535\\
090902B  & Band & 1.822 & 775 $\pm$ 11  &  -0.696 $\pm$ 0.012 &  -3.85  $\pm$ 0.31 & 374 $\pm$ 3 & 50-10000 & 9866\\
090926  & Band & 2.1062 & 314 $\pm$ 4  &   -0.75 $\pm$ 0.01 &   -2.59  $\pm$ 0.05 &  145 $\pm$ 4 & 8-1000 & 9933\\
091003  & Band & 0.8969 & 486.2  $\pm$ 23.6  &  -1.13 $\pm$ 0.01   &   -2.64  $\pm$ 0.24 &  37.6 $\pm$ 0.4 & 8-1000 & 9983\\
091020  & Band & 1.71 & 103 $\pm$ 68 &  -0.93 $\pm$ 0.6 & -1.9 $\pm$ 0.8 &  10.4 $\pm$ 2.1 & 20-2000 & 10057\\
091127  & Band & 0.49 & 36 $\pm$ 2  &   -1.27   $\pm$ 0.06    &   -2.20  $\pm$ 0.02 &  18.7 $\pm$ 0.2 & 8-1000 & 10204\\
091208B  & Band & 1.0633 & 124 $\pm$ 20.1  &   -1.44 $\pm$  0.07    &  -2.32 $\pm$ 0.47 &  5.8 $\pm$ 0.2 & 8-1000 & 10266\\
\hline %inserts single line
\end{tabular}
\\
\hspace{1mm}
References: \citep{Sakamoto}, \citep{McBreen}, \citep{Golenetskii}, \citep{Sakamoto2}, \citep{Bissaldi}, \citep{Bissaldi2}, \citep{Briggs}, \citep{Rau}, \citep{Golenetskii2}, \citep{Golenetskii3}, \citep{Barthelmy}, \citep{Wilson}, \citep{McBreen2}
\end{table*}

\section{The Cosmographic Amati relation}

At this point we can calculate the parameters of the Amati relation for the sample that we constructed previously.
This relation is a correlation of  type $ E_{iso} = a E_{p,i} ^{\gamma}$; however if we report it in a logarithmic basis, it reduces to the  form:
\begin{equation}
 \log_{10} E_{iso} =  A + \gamma \log_{10} E_{p,i}
\end{equation}
so we can report our sample in a diagram $ \log_{10} E_{iso} $ - $ \log_{10} E_{p,i} $ and perform a linear fit of the data, with weights given by the data errors on both the quantities involved.
An $ R^{2}$-test provides  an estimation of the reliability of the fit being $ R^2 = $ 0.772. This is   a good value, but not so suitable for our analysis.
The results of the fit, with errors amounting to a deviation of 3$\sigma$, and the corresponding covariance matrix are:

\begin{equation}
 A = 49.154 \pm 0.306 \quad \gamma = 1.444 \pm 0.117
\end{equation}

\begin{center}
 $\begin{Bmatrix}
  0.0136435 & -0.00509148 \\
 -0.00509148 & 0.00197731
\end{Bmatrix}$
\end{center}

A comparison with the results obtained by different interpolation methods  \citep{Liang}  shows a slight discrepancy between the parameters of the relation.
This fact  could be due to the calibration  in \citep{Liang}.  It depends  on the trend traced by SNeIa, while the cosmographic analysis takes into account the corrections due to physical parameters as  $q_0 $,  $ j_0 $.
Nevertheless the reason could  be another: since the SNeIa sample, used  here to calibrate the Amati relation, is different from that  in \citep{Liang}, where the authors adopted the catalog of 192 SNeIa  discussed in \citep{Wood}. This means that the slight difference in the results could be due to the different samples used for the calibration.

\begin{center}
\begin{figure*}
\includegraphics[width=16cm, height=9cm]{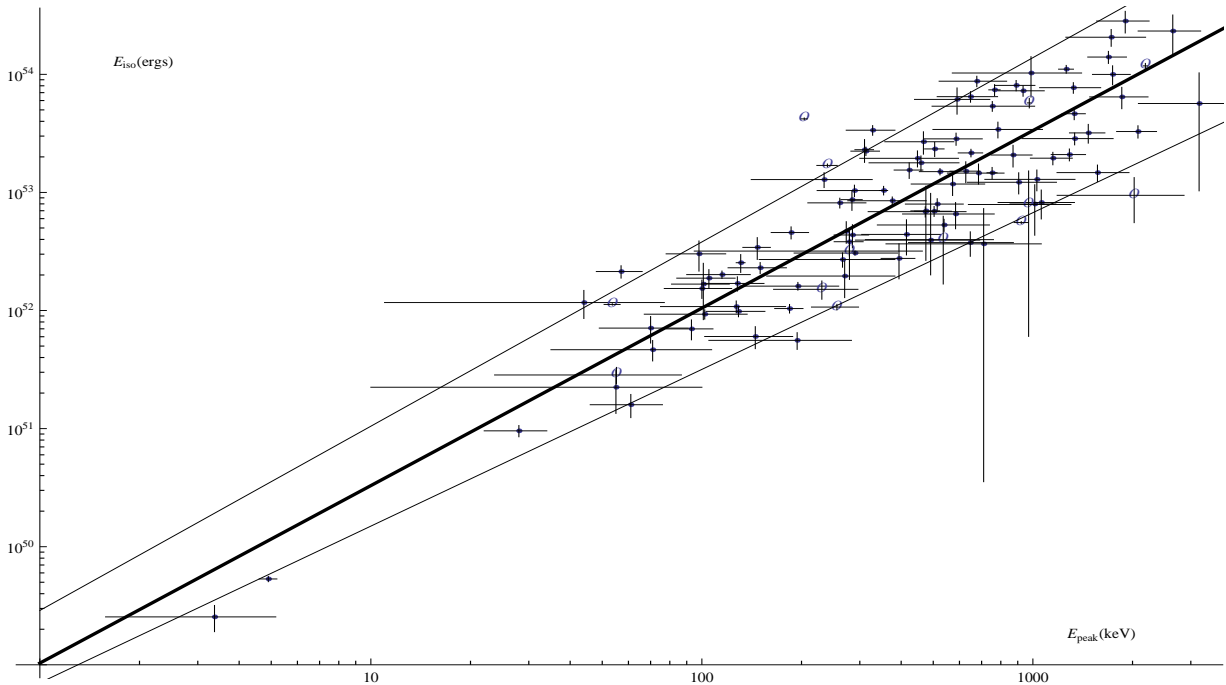}
\caption{Plot of the cosmographic Amati relation. The line of
prediction bounds represents a deviation of 3$\sigma$ from the
best fit line, the thick  line. The circle represents the 13 GRBs
added to the old sample, \citep{Amati09}. } \label{fig:no3}
\end{figure*}
\end{center}

In Fig.\ref{fig:no3},  it is shown the plot of the cosmographic Amati relation.
The confidence level curves are calculated as the 3$\sigma$ deviation from the best fit.
Note how the 13 GRBs added to the old sample, marked with a circle, are distributed about  the best-fit curve, indicating that the spectral analysis of these 13 GRBs is correct.

\begin{figure}
\includegraphics[width=9cm, height=6cm]{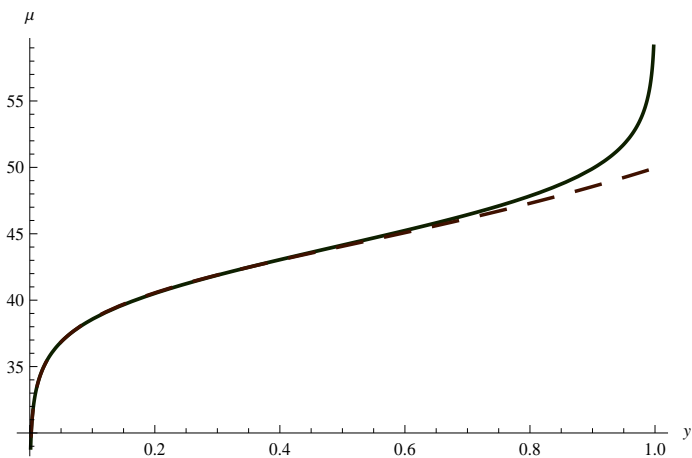}
\caption{Plot of  $\mu(y)$ computed for a fiducial $\Lambda$CDM
cosmological model, the continuous line, and for the reconstructed
$\mu(y)$ obtained by  the cosmographic fit of the SNeIa, the
dashed line. Note the slight deviation at very high redshift.}
\label{fig:no2}
\end{figure}

\section{Discussion and Conclusions}

The issue to extend the cosmic scale ladder up to medium-high
redshift  is an important questions of  modern cosmology. A
possible way to achieve this goal is to take into account GRBs,
the most powerful explosions in the Universe.  The energy emitted
by these objects spans about six orders of magnitude. However,
they cannot be assumed as standard candles in a proper sense.
Dispite of this lack, the existence of several correlations
between spectroscopic and photometric observable quantities of
GRBs allow us to solve in part this problem. The fundamental
pre-requisite to obtain such relations is to estimate
 the emitted  energy  in a way independent of the
cosmological model. In this paper,  we have considered a relation
for the luminosity distance $d_l$ that is independent on the
dynamics of the Universe, but, in principle, could work only at
small redshift. Although we have use a parameterization for the
redshift which allows  to transform the variable $z$ in a new
variable $y$, ranging in a  limited interval, we have seen that
the obtained luminosity distance at high-redshift differs slightly
from the fiducial model $ \Lambda$CDM at high redshift, see
Fig.\ref{fig:no3}. Nevertheless, since we obtained the curve $
d_l(y) $ by an analysis of the  SNeIa Union survey, that extends
up to a redshift of $ \sim $ 1.7, we achieved an independent
estimate at slightly higher redshift\footnote{ Estimates of the
Baryonic Acoustic Oscillations (BAO)  performed by forthcoming
surveys of clusters at intermediate redshift ($z \approx$ 2.5 -
3.5) will give better approximations of the curve $ d_l(y) $.}.
By the way using the $ d_l(y) $ obtained with the cosmographic fit
of the SNeIa, we have constrained a sample of GRBs in a
cosmology-independent way so that we have fitted  a cosmographic
Amati relation for GRBs. The results  are  similar to those
obtained from other analysis performed using other methods,
\citep{Schaefer,Liang,Amati}. It is important to stress the
independence from  cosmology and the calibration obtained by
SNeIa. In our opinion, this characteristic is relevant, from one
side, to constrain  cosmological models, in particular, dark
energy models, and, from another side, to check the physical
validity of the Amati  relation.

\paragraph{Acknowledgements.}

We warmly  thank L. Amati for providing us  the GRB sample  in \citep{Amati09}.
LI  warmly thanks also R. Benini for useful discussion and help with the Mathematica package data analysis.

\end{document}